\documentclass[a4paper,UKenglish,cleveref,autoref,thm-restate,authorcolumns]{lipics-v2019}

\title{The New Rewriting Engine of Dedukti}

\author{Gabriel Hondet}{Universit\'e Paris-Saclay, ENS
  Paris-Saclay, CNRS, Inria \and Laboratoire Sp\'ecification et
  V\'erification, 94235, Cachan, France}{}{}{}

\author{Frédéric Blanqui}{Universit\'e Paris-Saclay, ENS
  Paris-Saclay, CNRS, Inria \and Laboratoire Sp\'ecification et
  V\'erification, 94235, Cachan, France}{}{}{}

\authorrunning{G. Hondet and F. Blanqui}

\Copyright{Inria}



\category{System Description} 

\relatedversion{\url{https://hal.inria.fr/hal-02317471}}

\supplement{\url{https://github.com/deducteam/lambdapi.git}}

\acknowledgements{The authors thank Bruno Barras and Rodolphe Lepigre for their help in developing the new rewriting engine of Dedukti.}

\nolinenumbers 

\EventEditors{Zena M. Ariola}
\EventNoEds{1}
\EventLongTitle{5th International Conference on Formal Structures for Computation and Deduction (FSCD 2020)}
\EventShortTitle{FSCD 2020}
\EventAcronym{FSCD}
\EventYear{2020}
\EventDate{June 29--July 5, 2020}
\EventLocation{Paris, France}
\EventLogo{}
\SeriesVolume{167}
\ArticleNo{12}

\usepackage{fscd20}

\begin{document}

\maketitle

\begin{abstract}
    \dedukti{} is a type-checker for the $\lambda\Pi$-calculus modulo
  rewriting, an extension of Edinburgh's logical framework LF where
  functions and type symbols can be defined by rewrite rules. It
  therefore contains an engine for rewriting LF terms and types
  according to the rewrite rules given by the user. A key component
  of this engine is the matching algorithm to find which rules can be fired. In this paper, we
  describe the class of rewrite rules supported by \dedukti{} and
  the new implementation of the matching algorithm. \dedukti{}
  supports non-linear rewrite rules on terms with binders using
  higher-order pattern-matching as in Combinatory Reduction Systems
  (CRS). The new matching algorithm extends the technique of decision
  trees introduced by Luc Maranget in the \toolstyle{OCaml} compiler
  to this more general context.


\end{abstract}

\section{Introduction}\label{sec:introduction}

\dedukti{} is primarily a type-checker for the so-called
$\lambda\Pi$-calculus modulo rewriting, $\lambda\Pi/R$, an extension
of Edinburgh's logical framework LF \cite{harper93jacm} where function
and type symbols can be defined by rewrite rules. This means that
\dedukti{} takes as input type declarations and rewrite rules, and
check that expressions are well typed modulo these rewrite rules and
the $\beta$-reduction of $\lambda$-calculus.

The $\lambda\Pi$-calculus is the simplest type system on top of the
pure untyped $\lambda$-calculus combining both the usual simple types
of (functional) programming ({\em e.g.} the type
$\mathbb{N}\to\mathbb{N}$ of functions from natural numbers to
natural numbers) with value-dependent types ({\em e.g.} the type
$\tprod{n}{\mathbb{N}}{V(n)}$ of vectors of some given dimension). In
fact, a simple type $A\to B$ is just a particular case of
dependent type $\tprod{x}{A}{B}$ where $x$ does not occur in
$B$. Syntactically, this means that types are not defined prior to
terms as usual, but that terms and types are mutually defined.

Moreover, in $\lambda\Pi/R$, a term of type $A$ is also seen as a term
of type $B$ if $A$ and $B$ are equivalent not only modulo
$\beta$-reduction but also modulo some user-defined rewrite rules
$R$. Therefore, to check that a term $t$ is of type $A$, one has to be
able to check when two expressions are equivalent modulo
$\beta$-reduction and rewrite rules. This is why there is a
rewriting engine in \dedukti{}.

Thanks to the Curry-Howard correspondence between $\lambda$-terms and
proofs on the one hand, and (dependent) types and formulas on the
other hand, \dedukti{} can be used as a proof checker. Hence, in
recent years, many satellite tools have been developed in order to
translate to \dedukti{} proofs generated by automated or interactive
theorem provers: Krajono for Matita, Coqine for Coq, Holide of
OpenTheory (HOL Light, HOL4), Focalide for Focalize, Isabelle, Zenon,
iProverModulo, ArchSAT, etc. \cite{assaf19draft}.

By unplugging its type verification engine to only retain its
rewriting engine, \dedukti{} can also be used as a programming
language. Thanks to its rewriting capabilities, \dedukti{}
can be used to apply transformation rules on terms and formulas with
binders \cite{thire18lfmtp,cauderlier18itp}.

A rewrite rule is nothing but an oriented equation
\cite{baader98book}. Rewriting consists in applying some set of
rewrite rules $R$ (and $\beta$-reduction) as long as possible so as to
get a term in (weak head) normal form.  At every step it is therefore
necessary to check whether a term matches some left-hand side of a
rule of $R$. It is therefore important to have an efficient algorithm
to know whether a rule is applicable and select one:
\begin{example}\label{ex:match}
  Consider the following rules in the new Dedukti syntax (pattern
  variables are prefixed by {\tt\$} to avoid name clashes with other
  symbols):
\begin{lstlisting}
rule f (c (c $x)) a ↪ $x
with f       $x   b ↪ $x
\end{lstlisting}
To select the correct rule to rewrite a term, the naive algorithm matches the
term against each rule left-hand side from the top rule to the bottom one.
Let us apply the algorithm on the matching of the term
\lstinline{t = f (c (c e)) b}.

The first argument of \lstinline{t} is matched against the first argument of the
first left-hand side \lstinline{c (c $x)}. As \lstinline{c (c e)} matches
\lstinline{c (c $x)}, it succeeds. However, when we pass to the second argument,
\lstinline{b} does not match the pattern \lstinline{a}.
So the second rule is tried. Pattern \lstinline{$x} filters successfully
\lstinline{c (c e)}, and \lstinline{b} matches \lstinline{b}, so it succeeds.

Yet, matching \lstinline{c (c e)} against
\lstinline{c (c $x)} can be avoided.
Indeed, if we start by matching the second argument of \lstinline{f},
then the first rule is rejected in one comparison.
The only remaining work is to match \lstinline{c (c e)} against
\lstinline{$x}.
\end{example}

In \cite{maranget08ml}, Maranget introduces a domain-specific language
of so-called decision trees for describing matching algorithms, and a
procedure for compiling some set of rewrite rules into this
language. But his language and compilation procedure handle rewrite
systems whose left-hand sides are linear constructor patterns only. In
\dedukti{}, as we are going to see it soon, we use a more general
class of patterns containing defined symbols and
$\lambda$-abstractions. They can also be non-linear and contain
variable-occurrence conditions as in Klop's Combinartory Reduction
Systems (CRS) \cite{klop93tcs}.

In this paper, we describe an extension of Maranget's work to this
more general setting, and present some benchmark.

\subparagraph*{Outline of the paper}
In \autoref{sec-matching}, we start by
giving examples of the kind of rewrite rules that can be handled by
\dedukti{}, before giving a more formal definition. In
\autoref{sec-trees}, we present our extension of Maranget's decision
trees, their syntax and semantics, and how to compile a set of rewrite
rules into this language. In \autoref{sec-results}, we compare this
new implementation with previous ones and other tools implementing
rewriting. Finally, in \autoref{sec-conclu}, we discuss some
related work and conclude.


\section{Rewriting in \dedukti{}}
\label{sec-matching}

We will start by providing the reader with various examples of rewrite
rules accepted by \dedukti{} before giving a formal definition. To
this end, we will use the new \dedukti{} syntax (see
\url{https://github.com/Deducteam/lambdapi}). In this new syntax, one
can use Unicode characters, some function symbols can be written in
infix positions and, in rewrite rules, pattern variables need to be
prefixed by {\tt\$} to avoid name clashes with function symbols. Note
however that, for the sake of simplicity, we may omit some
declarations.

\dedukti{} can of course handle the ``Hello world!'' example of
first-order rewriting, the addition on unary natural numbers, as
follows:
\begin{lstlisting}
symbol ℕ: TYPE     symbol 0: ℕ     symbol s: ℕ → ℕ

symbol +: ℕ → ℕ → ℕ

rule     0  + $m ↪ $m
with (s $n) + $m ↪ s ($n + $m)
\end{lstlisting}

More interestingly is the fact that, in constrast to functional
programming languages like OCaml, Haskell or Coq, rule left-hand sides
can overlap each other. Consequently, in \dedukti{}, addition on unary
numbers can be more interestingly defined as follows:
\begin{lstlisting}
rule      0 + $m     ↪ $m
with (s $n) + $m     ↪ s ($n + $m)
with     $m + 0      ↪ $m
with     $m + (s $n) ↪ s ($m + $n)
\end{lstlisting}
With the first definition, one has $0+t$ equivalent to $t$ modulo
rewriting, written $0+t\simeq t$, for all terms $t$ (of type
$\mathbb{N}$), but not $t+0\simeq t$. Hence, the interest of the
second definition.

It is also possible to match on defined symbols and not just on
constructors like in usual functional programming languages. Hence,
for instance, one can add the following associativity rule on addition:
\begin{lstlisting}
rule ($x + $y) + $z ↪ $x + ($y + $z)
\end{lstlisting}

Moreover, one can use non-linear patterns, that is, require the
equality of some subterms to fire a rule like in:
\begin{lstlisting}
rule $x + (- $x) ↪ 0
\end{lstlisting}

Therefore, \dedukti{} can handle any first-order rewriting system
\cite{baader98book}. But it can also handle higher-order rewriting in
the style of Combinatory Reduction Systems (CRS) \cite{klop93tcs}.

The simplest example of higher-order rewriting is given by the {\tt
  map} function on lists, which applies an argument function to every
element of a list:
\begin{lstlisting}
symbol map: (ℕ → ℕ) → List → List

rule map $f (cons $x $l) ↪ cons ($f $x) (map $f $l)
\end{lstlisting}

Unlike first-order rewriting, function symbols can be partially
applied, including in patterns. Hence, in \dedukti{}, one can write
the following:
\begin{lstlisting}
symbol id: ℕ → ℕ
rule id $x ↪ $x
rule plus 0 ↪ id with plus (s $n) $m ↪ s (plus $n $m)
rule map id $l ↪ $l
\end{lstlisting}

It is also possible to match $\lambda$-abstractions as follows:
\begin{lstlisting}
symbol cos: ℝ → ℝ
symbol sin: ℝ → ℝ
symbol *: (ℝ → ℝ) → (ℝ → ℝ) → (ℝ → ℝ)
symbol diff: (ℝ → ℝ) → (ℝ → ℝ)

rule diff(λx,sin($v[x])) ↪ diff(λx,$v[x]) * cos
\end{lstlisting}

Following the definition of CRS, in a rule left-hand side, a
higher-order pattern variable can only be applied to distinct bound
variables (this condition could be slightly relaxed though
\cite{libal16fscd}). A similar condition appears in $\lambda$Prolog
\cite{miller91jlc}. It ensures the decidability of matching.

It can also be used to check variable-occurrence conditions.
The differential of a constant function can thus be simply defined as
follows in \dedukti{}:
\begin{lstlisting}
rule diff(λx,$v) ↪ λx,0
\end{lstlisting}

While in the rule for {\tt sin}, we had \(\patt[\mathtt{v}][x]\),
meaning that the term matching \(\patt[\mathtt{v}][x]\) may depend on {\tt x},
here we have \(\patt[\mathtt{v}]\) applied to no bound variables,
meaning that the term matching \(\patt[\mathtt{v}]\) cannot depend on \(x\).

\subsection{Terms, Patterns, Rewrite Rules and Matching, Formally}

We now define more formally terms, patterns, rewrite rules and
rewriting. Following \cite{barendregt92chapter}, the terms of the
$\lambda\Pi$-calculus are inductively defined as follows:
\begin{equation*}
  t,u ::= \mathtt{TYPE}\mid \mathtt{KIND}\mid x\mid f\mid tu\mid
  \tabs{x}[t]{u}\mid \tprod{x}{t}{u}
\end{equation*}

\noindent
where $x$ is a term variable, $f$ is a function symbol, $tu$ is the
application of the function $t$ to the term $u$, $\tabs{x}[t]{u}$ is
the function mapping $x$ of type $t$ to $u$, which type is the
dependent product $\tprod{x}{t}{u}$. The simple type $t\to u$
is syntactic sugar for $\tprod{x}{t}{u}$ where $x$ is any fresh term
variable not occurring in $u$.

In $\tabs{x}[t]{u}$ and $\tprod{x}{t}{u}$, the occurrences of $x$ in
$u$ are bound, and terms equivalent modulo renaming of their bound
variables are identified, as usual. In \dedukti{}, this is implemented
by using the Bindlib library \cite{lepigre18lfmtp}.

A (possibly empty) ordered sequence of terms $t_1,\ldots,t_n$ is
written $\mvec{t}$ for short.

Patterns are inductively defined as follows:
\begin{equation*}
  p ::= \patt[x][\mvec{y}] \mid f\mvec{p}\mid \tabs{y}{p}
\end{equation*}
where $\patt[x]$ is a pattern variable and $\mvec{y}$ is a sequence of
distinct bound variables.

A rewrite rule is a pair of terms, written $\ell\to r$,
such that $\ell$ is
a pattern of the form $f\mvec{p}$ and every pattern variable occurring
in $r$ also occurs in $\ell$.

In the following, we will assume given a set of user-defined rewrite
rules $R$.

Matching a term $t$ against a pattern $p$ whose bound variables are in
the set $V$, written \( p \filters[V] t\) is inductively defined as
follows:
\begin{subequations}\label{eq:instance-def}
  \begin{align}
    \patt[x][\mvec{y}] &\filters[V] t
    && \text{iff } \freevar(t) \cap V \subseteq \{\mvec{y}\}
    \tag{MatchFv}\label{eq:InstFv} \\
       \rwsymbol{f}\, p_1\, \dots\, p_n
    &\filters[V] \rwsymbol{f}\, t_1\, \dots\, t_n
       && \text{iff } (p_1\, \dots\, p_n) \filters[V] (t_1\, \dots\, t_n)
       \tag{MatchSymb}\label{eq:InstSymb} \\
    \lambda y,p &\filters[V] \lambda y:A,t
    && \text{iff } p \filters[V\uplus\{y\}] t
    \tag{MatchAbst}\label{eq:InstAbst}\\
       (p_1\ \dots\ p_n) &\filters[V] (t_1\ \dots\ t_n)
    && \text{iff } \forall i, p_i \filters[V] t_i
    \wedge \forall j, p_i = p_j\Rightarrow t_i = t_j
       \tag{MatchTuple}\label{eq:InstTuple}
  \end{align}
\end{subequations}
and we say that the term $t$ \emph{matches} the pattern $p$ or that the pattern
$p$ \emph{filters} the term $t$.

The indexing set \( V \) of variables
is used to record which binders have been traversed,
which is necessary to perform variable-occurrence tests.

The condition in the (\ref{eq:InstTuple}) rule translates non-linearity
conditions: if a variable occurs twice in a pattern, then the matching
values must be equal.


\section{Implementing Matching With Decision Trees}
\label{sec-trees}

The rewriting engine described in this paper is based on the work of
Maranget \cite{maranget08ml}. Maranget introduces a domain-specific
language for matching and an algorithm to transform a (ordered) list
of first-order linear constructor patterns into a program in this
language. In this section, we explain how we extend Maranget's
language and compilation procedure to our more general setting with
non-linear higher-order patterns, partially applied function symbols,
and no order on patterns.

We start by defining the language of decision trees \( D \) and switch
case lists \(L\):
\begin{equation*}
  \begin{gathered}
    \begin{array}{rcl}
      D, E & ::= & \trleaf(r) \mid \trfail \mid \trswap_i(D) \mid
                   \trstore(D) \mid \trswitch(L) \\
        & \phantom{::}| & \trbinNl(D, \{i, j\}, E) \mid \trbinCl(D, (n, X), E)\\
      L & ::= & (s, D) \liCons L \mid (\lambda, D) \liCons L_{\lambda} \mid T\\
      L_{\lambda} & ::= & (s, D) \liCons L_{\lambda} \mid T\\
      T & ::= & (*, D) \liCons \liNil \mid \liNil
    \end{array}\\
  \end{gathered}
\end{equation*}
where \( r \) is a rule right-hand side, \( i, j \)
and \( n \) are integers, \( X \) is a finite set of variables.
For case lists, \( s \) is a function symbol annotated with the number of
arguments it is applied to,
\( \liCons \) is the cons operator on lists and \( \liNil \)
is the empty list.

An element of a switch case list is a pair mapping:
\begin{itemize}
\item a function symbol $s$ to a tree for matching its arguments,
\item a \( \lambda \) to a tree for matching the body of an abstraction,
\item a default case \( * \) to a tree for matching the other arguments.
\end{itemize}
Note that a list $L_\lambda$ has no element \( (\lambda, D) \) and,
in a list $L$, there is at most one element of the form \( (\lambda, D) \).
Finally, in both cases, there is at most one element of the form \( (*, D) \)
and, if so, it is the last one (default case).

\subparagraph*{Semantics}
Decision trees are evaluated along with a stack of terms \( \mvec{v}
\) to filter and an array \( \mvec{s} \) used by the decision
tree to store elements.
Informally, the semantics of each tree constructor is as follows:
\begin{description}
\item[$\trleaf(r)$] matching succeeds and yields right-hand side \( r
  \).
\item[$\trfail$] matching fails.
\item[$\trswap_i(D)$] moves the $i$th element of \( \mvec{v} \) to the
  top of $\mvec{v}$ and carries on with \( D \).
\item[$\trstore(D)$] stores the top of the stack into \(
  \mvec{s} \) and continues with \( D \).
\item[$\trswitch(L)$] branches on a tree in \( L \) depending on the
  term on top of \( \mvec{v} \).
\item[$\trbinNl(D, \{i, j\}, E)$] checks whether $s_i$ and s$_j$ are
  equal and continues with \( D \) if this is the case, and
  \( E \) otherwise.
\item[$\trbinCl(D, (n, X), E)$] checks whether $\freevar(s_n)\subseteq X$ and
  continues with \( D \) if this is the case, and with \( E \)
  otherwise.
\end{description}

\begin{example}\label{ex:tree}
  The matching algorithm described in \autoref{ex:match} can be
  represented by the following decision tree:
\begin{equation*}
  \begin{aligned}
  \trswap_2(\trswitch(
    [
      &(
        \rws{a}_0,
        \trswitch([(\rws{c}_0,
          \trswitch([(\rws{c}_0,
            \trleaf(\patt[x]))]))]));\\
      &(
        \rws{b}_{0},
        \trleaf(\patt[x])
        )
    ])
)
  \end{aligned}
\end{equation*}
which can be graphically represented as follows:
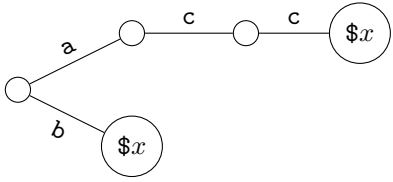
\begin{figure}[h]
\begin{tikzpicture}[sloped, grow=right]
  \node[circle,draw] {}
  child{
    node[circle,draw]{\patt[x]}
    edge from parent node[below]{\texttt{b}}
  }
  child{
    node[circle,draw]{}
    child{
      node[circle,draw]{}
      child{
        node[circle,draw]{\patt[x]}
        edge from parent node[above]{\texttt{c}}
      }
      edge from parent node[above]{\texttt{c}}
    }
    edge from parent node[above]{\texttt{a}}
  };
\end{tikzpicture}
\caption{Graphical representation of the decision tree of \autoref{ex:tree}}\label{fig:dec-tree-intro}
\end{figure}
where leaves are the right-hand sides of the rewrite rules and
a path from the root to a leaf is a successful matching.
The tree of \autoref{fig:dec-tree-intro} can be used to rewrite any term
of the form \( \mathtt{f}\ \mvec{t} \).
The sequence of operations to filter the term
\( \rws{f}\ (\rws{f}\ \rws{a})\ \rws{b} \) can be read from the tree.
The initial vector \( \mvec{v} \) is
\( \mvec{v} = (\rws{f}\ \rws{a}, \rws{b}) \)
and the array \( \mvec{s} \) won't be necessary here.
\begin{enumerate}
  \item The \( \trswap_2 \) transforms \( \mvec{v} \) into
    \( (\rws{b}, \rws{f}\ \rws{a}) \),
    so the next operations will be carried out on \( \rws{b} \).
\item The \( \trswitch \) node with the case list
  \( [(\rws{a}_0, D),(\rws{b}_0, E)] \) allows to branch on \( D \)
  or \( E \) depending on the term on top of \( \mvec{v} \), that is,
  \( \rws{b} \). Since \( \rws{b} \) is applied to no argument, it
  matches \( \rws{b}_0 \) and filtering continues on \( E \).
  The stack is now \( \mvec{v} = (\rws{f}\ \rws{a}) \).
\item Node \( E \) is in fact a \( \trleaf \) and so the matching succeeds.
\end{enumerate}
Note that the top symbol \(\rws{f}\) is not matched.
Top symbols are analysed prior to filtering as they are needed to get the
appropriate decision tree to filter the arguments.
\end{example}

The formal semantics is given in \autoref{fig:eval-dtree}.
An evaluation is written as a judgement
\( \mvec{v}, \mvec{s}, V \vdash D \rightsquigarrow r \) which can be read:
``stack \( \mvec{v} \), store $\mvec{s}$ and abstracted variables $V$
yield the term \( r \) when matched against tree \( D \)''.
We overload the comma notation, using it for the cons \((\rws{s}, \mvec{v})\)
and the concatenation \((\mvec{v}, \mvec{w})\). The \(|\) is used as the
alternative.

Matching succeeds with the \textsc{Match} rule.
Terms are memorised on the stack \(\mvec{s}\)
using the \textsc{Store} rule.
Matching on a symbol is performed with the \textsc{SwitchSymb} rule.
If the stack has a term \(\rws{f}\) applied on top and
the switch-case list \(L\) contains an element \((\rws{f}, D)\),
then the symbol \(\rws{f}\) can be removed, and matching continues using
sub-tree \(D\).
The rule \textsc{SwitchDefault} allows to match on any symbol or
abstraction, provided that the switch-case list \(L\) has a default case
(and that we can apply neither rule
\textsc{SwitchSymb} nor \textsc{SwitchAbst}).
The binary constraint rules guide the matching depending on failure or success
of the constraints.
The last three rules allow to search for a symbol in a switch-case list.
A judgement \(s \vdash L \rightsquigarrow p\) reads ``looking for symbol \(s\)
in list \(L\) yields pair \(p\)''.
\textsc{Cont} skips a cell of the list,
\textsc{Default} returns unconditionally the default cell of the
list (which is the last by construction) and
\textsc{Found} returns the cell that matches the symbol looked for.

\begin{minfrules}{Evaluation of decision trees}{fig:eval-dtree}
  \inferrule[Match]{ }{\mvec{v}; \mvec{s}; V \vdash \trleaf(k)
    \rightsquigarrow k}
  \and
  \inferrule[Swap]{(v_{i}, \dots, v_{1},
    \dots, v_n); \mvec{s}; V \vdash D
    \rightsquigarrow k
  }{%
    (v_1, \dots, v_{i}, \dots v_n); \mvec{s}; V \vdash
    \trswap_i(D) \rightsquigarrow k }
  \and
  \inferrule[Store]{%
    \mvec{v}; \mvec{s}\,v_{1}; V
    \vdash
    D \rightsquigarrow k
  }{%
    \mvec{v}; \mvec{s}; V \vdash
    \trstore(D) \rightsquigarrow k
  }\label{infrule:store}
  \and
  \inferrule[SwitchSymb]{%
    \rwsymbol{f} \vdash L \rightsquigarrow
    (\rwsymbol{f}, D) \\
    (\mvec{w}, \mvec{v}); \mvec{s}; V
    \vdash
    D \rightsquigarrow k
  }{%
    (\rwsymbol{f}\ \mvec{w}, \mvec{v});
    \mvec{s}; V \vdash \trswitch(L) \rightsquigarrow
    k
  }
  \and
  \inferrule[SwitchDefault]{%
    (\rwsymbol{s}_{\ell} \mid \lambda) \vdash L \rightsquigarrow
    (*, D) \\
    \mvec{v}; \mvec{s}; V \vdash D
    \rightsquigarrow k
  }{%
    ((\rwsymbol{s}\, w_1\, \cdots\, w_\ell \mid \tabs{x}{w}), \mvec{v});
    \mvec{s}; V \vdash \trswitch(L)
    \rightsquigarrow k
  }
  \and
  \inferrule[SwitchAbst]{%
    \lambda \vdash L
    \rightsquigarrow (\lambda, D) \\
    (w, \mvec{v});
    \mvec{s}; V \cup \{ x \}
    \vdash D
    \rightsquigarrow k
  }{%
    (\tabs{x}{w}, \mvec{v}); \mvec{s}; V
    \vdash
    \trswitch(L) \rightsquigarrow k
  }
  \and
  \inferrule[BinClSucc]{%
    \freevar(s_i)\cap X\subseteq V\\
    \mvec{v}; \mvec{s}; V
    \vdash D \rightsquigarrow k
  }{%
    \mvec{v}; \mvec{s}; V \vdash
    \trbinCl(D, (i, X), E)
    \rightsquigarrow k }
  \and
  \inferrule[BinClFail]{%
    \freevar(s_i)\cap X\not\subseteq V\\
    \mvec{v}; \mvec{s}; V \vdash E \rightsquigarrow k
  }{%
    \mvec{v}; \mvec{s}; V \vdash
    \trbinCl(D, (i,X), E) \rightsquigarrow k }
  \and
  \inferrule[BinNlSucc]{%
    s_j=s_j\\
    \mvec{v}; \mvec{s}; V \vdash D \rightsquigarrow k
  }{%
    \mvec{v}; \mvec{s}; V
    \vdash \trbinNl(D, \{i, j\}, E)
    \rightsquigarrow k }
  \and
  \inferrule[BinNlFail]{%
    s_i\neq s_j\\
    \mvec{v}; \mvec{s}; V \vdash E \rightsquigarrow k
  }{%
    \mvec{v}; \mvec{s}; V \vdash \trbinNl(D, \{i, j\}, E)
    \rightsquigarrow k }
  \and
  \inferrule[Found]{ }{s \vdash
    (s, D) :: L
    \rightsquigarrow (s, D)}
  \and
  \inferrule[Default]{ }{s \vdash (*, D)
    \rightsquigarrow (*, D)}
  \and
  \inferrule[Cont]{%
    s \ne s' \\ s \vdash L
    \rightsquigarrow (s|*, D)
  }{%
    s \vdash (s', D) ::
    L
    \rightsquigarrow (s|*, D)
  }
\end{minfrules}

\subsection{Matrix Representation of Rewrite Systems}\label{sec:matrix}

In order to compile a set of rewrite rules into this language, it is
convenient to represent rewrite systems as tuples containing a matrix
and three vectors.
The matrix contains the patterns and can have lines of different lengths
because function symbols can be partially applied.
The vectors contain the right-hand side of the rewriting system and the
constraints.
Hence, a rewrite system for a function symbol $f$, that is,
a set of rewrite rules $f\mvec{p}^1\to r^1,\ldots,f\mvec{p}^m\to r^m$
is represented by:
\begin{equation*}
  \pattMat{%
    \begin{bmatrix}
      p_1^1 & \cdots & p_{n_1}^1\\
      p_1^2 & \cdots & p_{n_2}^2\\
      & \vdots & \\
      p_1^m & \cdots & p_{n_m}^m
    \end{bmatrix}
  }{%
    \begin{bmatrix}
      N_1 \\ N_2 \\ \vdots \\ N_m
    \end{bmatrix}
  }{%
    \begin{bmatrix}
      C_1 \\ C_2 \\ \vdots \\ C_m
    \end{bmatrix}
  }{%
    \begin{bmatrix}
      r_1 \\ r_2 \\ \vdots \\ r_m
    \end{bmatrix}
  }
\end{equation*}
where \( C_i \) encodes the variable-occurrence constraints in \(
\mvec{p}^i \) and \( N_i \) encodes non-linearity constraints in \(
\mvec{p}^i \).

A variable-occurrence constraint given by a pattern variable \(
\patt[x][\mvec{y}] \) is encoded as a pair \( (a, \mvec{y}) \) where
\( a \) is the position of the variable in the main term.

Non-linearity constraints between two terms at positions
\( a \) and \( b \) are encoded by the unordered pair \( \{ a, b \} \).

In the above matrix, we can then replace a pattern of the form
\( \patt[x][\mvec{y}] \) or \( \patt[x] \) by \( \patt \).

For the sake of completeness, we recall the definition of positions:

\begin{definition}[Positions in a term]\label{defn:position}
The set of \emph{positions}
    of a term \( t \) is the set of words
    over the alphabet of positive
    integers inductively defined as follows:
    \begin{itemize}
    \item \( \posset(x) \mdefn \{ \epsilon \} \)
    \item
      \(
        \posset(\rwsymbol{f}\ t_1\ \cdots\ t_n) \mdefn \{ \epsilon \} \cup \bigcup_{i = 1}^{n}
        \{ ia \mid a \in \posset(t_i) \}
      \)
    \item \( \posset(\tabs{x}{t}) \mdefn \{ \epsilon \} \cup \{ 1a \mid
      a \in \posset(t) \} \)
    \end{itemize}
    The position \( \epsilon \) is called
    the \emph{root position} of the term \( t \) and the symbol
    at this position is called the \emph{root symbol} of
    \( t \).

 For \( a \in \posset(t) \), the \emph{subterm of \( t \) at
      position \( a \)}, denoted by \( \msubterm{t}{a} \), is defined
    by induction on the length of \( a \),
      \( \msubterm{t}{\epsilon} \mdefn t\) and
      \( \msubterm{\rwsymbol{f}\ t_1\ \cdots\ t_n}{ia} \mdefn
      \msubterm{t_i}{a} \)

          The notion of position is extended to sequences of terms by taking
\( \msubterm{\mvec{t}}{ia} \mdefn \msubterm{t_i}{a} \).
\end{definition}

\begin{example}
  The rewrite system
  \begin{lstlisting}[language=Dedukti]
rule f a  (`$\lambda$`x,`$\lambda$`y,$g[x]) `$\hookrightarrow$` 0
with f $x $x            `$\hookrightarrow$` 1
with f a  b             `$\hookrightarrow$` 2
  \end{lstlisting}
  is represented by the following matrix:
  \begin{equation*}
    \pattMat{%
      \begin{bmatrix}
        \rwsymbol{a} & \tabs{x}{\tabs{y}{\patt}} \\
        \patt & \patt \\
        \rwsymbol{a} & \rwsymbol{b}
      \end{bmatrix}
    }{%
      \begin{bmatrix}
        \emptyset \\ \{ \{ 1, 2 \} \} \\ \emptyset
      \end{bmatrix}
    }{%
      \begin{bmatrix}
        \{ (211, (x)) \} \\ \emptyset \\ \emptyset
      \end{bmatrix}
    }{%
      \begin{bmatrix}
        0 \\ 1 \\ 2
      \end{bmatrix}
    }.
  \end{equation*}
  The variable-occurence constraint of the first rule is encoded by \(
  (211, (x)) \) since only the variable \( x \) is authorised in \(\patt[g][x]\).
  The non-linearity constraint \lstinline{f &x &x} is translated by
  \( \{ 1, 2 \} \), hence the constraints set \( \{\{ 1, 2 \}\} \).
\end{example}

\subsection{Compiling Rewrite Systems to Decision Trees}\label{sec:comp-proc}

We will describe the compilation process as a non-deterministic
recursively defined relation \( \rhd \) between matrices and decision
trees.

To this end, we use the transformations on matrices
defined in \autoref{tab:decomp}.
\begin{itemize}
\item \( \specialise\left(\rws{f}, a, \pattMat{P}{\mvec{N}}{\mvec{C}}{\mvec{r}}\right) \)
  keeps rows whose first pattern
  filters the application of function \(\rws{f}\) \(a\) arguments:
  \begin{example}\label{ex:specialise}
    Let \( P =
          \begin{bmatrix}
            \rws{r}\ \patt[x] & \rws{q}\\
            \rws{r}\ & \rws{f}\ \patt[x]\\
            \patt[x] & \rws{r}\\
            \tabs{x}{\patt[x]} & \tabs{x}{\rws{r}}
          \end{bmatrix} \). Then,
    \[
      \specialise\left(\rws{r}, 1,
        \pattMat{P}{\mvec{N}}{\mvec{C}}{\mvec{r}}\right) =
        \pattMat{%
          \begin{bmatrix}
            \patt[x] & \rws{q}\\
            \patt & \rws{r}
          \end{bmatrix}
        }{\mvec{N}}{\mvec{C}}{\begin{bmatrix}r_1 \\ r_3 \end{bmatrix}}
    \]
  \end{example}
\item \( \speclam\left(\pattMat{P}{\mvec{N}}{\mvec{C}}{\mvec{r}}\right) \)
  keeps rows whose first pattern filters a $\lambda$-abstraction:
  \begin{example} Let \( P \) be the same as in \autoref{ex:specialise}.
    \[
      \speclam
        \pattMat{P}{\mvec{N}}{\mvec{C}}{\mvec{r}} =
        \pattMat{%
          \begin{bmatrix}
            \patt & \rws{r}\\
            \patt[x][x] & \tabs{x}{\rws{r}}
          \end{bmatrix}
        }{\mvec{N}}{\mvec{C}}{\begin{bmatrix}r_3 \\ r_4\end{bmatrix}}
    \]
  \end{example}
\item \( \default\pattMat{P}{\mvec{N}}{\mvec{C}}{\mvec{r}} \)
  keeps rows whose first pattern is a pattern variable:
  \begin{example} Let \( P \) be the same as in \autoref{ex:specialise}.
    \[
      \default
        \pattMat{P}{\mvec{N}}{\mvec{C}}{\mvec{r}} =
        \pattMat{%
          \begin{bmatrix}
            \rws{r}
          \end{bmatrix}
        }{\mvec{N}}{\mvec{C}}{\begin{bmatrix} r_3 \end{bmatrix}}
    \]
  \end{example}
\end{itemize}
To sum up, given a pattern matrix \( P \),
a simplification function removes rows
of \( P \) that are not compatible with some assumption on the form of
the first pattern.

\begin{table}
  \centering
  \begin{tabularx}{1.0\linewidth}{c c c c}
    \toprule
    Pattern \( p_1^j \)
    & Rows of \( \specialise(\rwsymbol{f}, a, P \to A) \)
    & Rows of \( \speclam(P \to A) \)
    & Rows of \( \default(P \to A) \) \\
    \midrule
    \( \rwsymbol{f}\ q_1\ \cdots\ q_a \)
    & \( q_1\ \cdots\ q_a\ p_2^j\ \cdots p_n^j \) & No row & No row \\
    \( \rwsymbol{f}\ q_1\ \cdots\ q_b \) & No row if \( a \ne b \) & No row & No row \\
    \( \rwsymbol{f}\ q_1\ \cdots\ q_b \) & No row & No row & No row\\
    \( \tabs{x}{q} \) & No row
    & \( q\ p_2^j\ \cdots p_n^j \) & No row \\
    \( \_ \) & \( \overbrace{\_\ \cdots\ \_}^{\times a} \)
    & \( \_\ p_2^j\ \cdots\ p_n^j \) & \( p_2^j\ \cdots\ p_n^j \) \\
    \bottomrule
  \end{tabularx}
  \caption{Decomposition operators}\label{tab:decomp}
\end{table}

The same idea is used for constraints.
Note that we will abuse set notations and write \(k \in N\)
or \(N \backslash \{k\}\)
even if \(N\) is not a set of elements of the type of \(k\).
In that case \(k \in N\) is false and \(N \backslash \{k\}\) is \(N\).
\begin{itemize}
\item \( \condsucc(k, (P,\mvec{N}, \mvec{C}, \mvec{r})) \)
  keeps all the rows and simplify the constraint sets
  \[ \condsucc(k, \pattMat{\mvec{p}}{N}{C}{r}) \mdefn
    \pattMat{\mvec{p}}{N \backslash \{k\}}{C \backslash \{k\}}{r} \]
\item \( \condfail(k, (P, \mvec{N}, \mvec{C}, \mvec{r})) \) keeps rows that
  don't have \( k \) in their constraint sets
  \[
  \condfail(k, \pattMat{\mvec{p}}{N}{C}{r}) \mdefn
  \begin{cases}
    \text{No row} & \text{if } k \in N \text{ or } k \in C \\
    \pattMat{\mvec{p}}{N}{C}{r} & \text{if } k \not\in N
                                  \text{ and } k \not\in C
  \end{cases}
  \]
\end{itemize}

A compilation process consists in reducing the matrix step by step,
compiling the sub-matrices and aggregating the sub-trees obtained
using the node that corresponds to the computed sub-matrices
(e.g.\ a \( \trswitch \) if the \( \specialise \), \( \default \) and
\( \speclam \) sub-matrices have been computed).

To say that the matrix \( \pattMat{P}{\mvec{N}}{\mvec{C}}{\mvec{r}} \)
compiles to the decision tree \( D \), we write
$$\left(
  \mvec{\rho}, \pattMat{P}{\mvec{N}}{\mvec{C}}{\mvec{r}}, n, \mathcal{E}
  \right) \rhd D $$
where:
\begin{itemize}
\item \( \mvec{\rho} \) are the positions in the term that will be
  matched against during evaluation.
\item \( \mathcal{E} \) is a map from positions to integers such that
  \( \mathcal{E}(\rho) \) is the index in \( \mvec{s} \) of the subterm at position $\rho$ used during the evaluation of decision trees.
  The empty map is denoted \( \emptyset \).
\item $n$ is the size of the store, which is incremented each time
  an element is added.
\end{itemize}

We now describe the compilation process implemented in \dedukti{}:

\begin{definition}[Compilation]
\begin{enumerate}
\item\label{enum:tc-empty}
  If the matrix \( P \) has no row (\( m = 0 \)), then matching always
  fails, since there is no rule to match,
  \begin{equation}
    \label{eq:treecomp-fail}
    \mvec{\rho}, \pattMat{\emptyset}{\mvec{N}}{\mvec{C}}{\mvec{r}}, n, \mathcal{E}
    \rhd \trfail
  \end{equation}
\item\label{enum:tc-vars}
  If there is a row \( k \) in \( P \) composed of unconstrained
  variables, matching succeeds and yields right-hand side \( r \)
  \begin{equation}
    \label{eq:treecomp-yield}
    \left(\mvec{\rho},
    \begin{pmatrix}
      p_1^1 & \cdots & p_{n_{1}}^1 & N_1 & C_1 & \to & r_{1} \\
      & & & \vdots \\
      \patt & \cdots & \patt & \emptyset & \emptyset & \to & r_k \\
      & & & \vdots \\
      p_1^m & \cdots & p_{n_m}^m & N_m & C_m & \to & r_m 
    \end{pmatrix}, n, \mathcal{E}
    \right) \rhd \trleaf(r_k)
  \end{equation}
\item\label{enum:tc-recurse}
  Otherwise, there is at least one row with either a symbol or a
  constraint or an abstraction.  We can choose to either specialise on
  a column or solve a constraint.
  \begin{enumerate}
  \item\label{enum:tc-rec-spec}
    Consider a specialisation on the first column, assuming it
    contains at least a symbol or an abstraction.

    If \( \rho_1 \) is constrained in some \( N_i \) or \( C_i \),
    then define \( n' = n + 1 \) and
    \( \mathcal{E}' = \mathcal{E} \cup \{ \rho_1 \mapsto n \} \).
    Otherwise, let \( n' = n \) and \( \mathcal{E}' = \mathcal{E} \).

    Let \( \Sigma \) be the set of root symbols of the terms of the
    first column and \( k \) the number of arguments
    \( f \) is applied to.
    Then for each \(f \in \Sigma \), we compile
    \begin{equation*}
      \label{eq:treecomp-symbol}
      \left(
        (\msubterm{\rho_1}{1}\; \cdots\;
        \msubterm{\rho_1}{k}\ \rho_2\;
        \cdots\; \rho_n),
        \specialise\left(
          f, k, \pattMat{P}{\mvec{N}}{\mvec{C}}{\mvec{r}}
        \right), n', \mathcal{E}'
        \right) \rhd D_{f_{k}}
    \end{equation*}

    Let \( L \) be the switch case list defined as
    (we use the bracket notation for list comprehension as
    the order is not important here)
    \begin{equation*}
      \label{eq:treecomp-case-mapping}
      L \mdefn \left[ (f, D_{f_{k}}) | f \in \Sigma \right]
    \end{equation*}

    If there is an abstraction in the column,
    the \( \speclam \) sub-matrix is computed and compiled to \(
    D_{\lambda} \), and an abstraction case is added to the mapping
    \begin{equation*}
      \label{eq:treecomp-abstraction}
      \begin{gathered}
        \left(
          (\msubterm{\rho_1}{1}\; \rho_{2}\; \cdots\; \rho_n),
          \speclam\left(
            \pattMat{P}{\mvec{N}}{\mvec{C}}{\mvec{r}}
          \right), n', \mathcal{E}'
        \right) \rhd D_\lambda \\
        L \mdefn
        [ (s, D_{f_{k}}) \mid f \in \Sigma ] \liCons (\lambda,
        D_{\lambda}) \liCons \liNil
      \end{gathered}
    \end{equation*}

    Similarly, if the column contains a variable,
    the \( \default \) sub-matrix is computed and compiled to \( D_{*} \),
    and  the mapping is completed with a
    default case, (the abstraction case may or may not be present)
    \begin{equation*}
      \label{eq:treecomp-default}
      \begin{gathered}[t]
        \left((\rho_2\; \cdots\; \rho_n),
          \default\left(
            \pattMat{P}{\mvec{N}}{\mvec{C}}{\mvec{r}}
            \right), n', \mathcal{E}' \right) \rhd D_*\\
          L \mdefn [ (f_{k}, D_{f_{k}}) | s \in \Sigma ] \liCons
          (\lambda, D_{\lambda}) \liCons (*, D_*) \liCons \liNil
      \end{gathered}
    \end{equation*}

    Now that the switch case list \( L \) is complete (all the symbols,
    the abstractions and the pattern variables are handled) and the
    sub-trees are defined and related to their pattern matrix,
    we can create the top node \( \trswitch(L) \).

    Furthermore, if \( \rho_1 \) is constrained, the term must be
    saved during evaluation. In that case, we add a \( \trstore \) node,
    \begin{equation*}
      \left(
        \mvec{\rho}, \pattMat{P}{\mvec{N}}{\mvec{C}}{\mvec{r}}, n,
        \mathcal{E}
      \right) \rhd \trstore(\trswitch(L)).
    \end{equation*}
    Otherwise,
    \begin{equation*}
      \left(
        \mvec{\rho}, \pattMat{P}{\mvec{N}}{\mvec{C}}{\mvec{r}}, n,
        \mathcal{E}
      \right) \rhd \trswitch(L).
    \end{equation*}

  \item\label{enum:tc-rec-closedness} If a term has been stored and is
    subject to a closedness constraint, then this constraint can be
    checked.

    That is, for any position \( \mu \) such that \(
    \mathcal{E}(\mu) \) is defined and there is a constraint set \(
    C_i \) and a variable set \( V \) such that \( (\mu, V) \in C_i
    \),
    we compute the sub-matrices \( \condsucc \) and \(
    \condfail \) and we compile them to \( D_{s} \) and \( D_f \),
    \begin{equation*}
      \left(
        \mvec{\rho}, \condsucc\left((\mu, V),
          \pattMat{P}{\mvec{N}}{\mvec{C}}{\mvec{r}}
        \right), n,
        \mathcal{E}
        \right) \rhd D_s
    \end{equation*}
    \begin{equation*}
      \left(
        \mvec{\rho}, \condfail\left((\mu, V),
          \pattMat{P}{\mvec{N}}{\mvec{C}}{\mvec{r}} 
        \right), n,
        \mathcal{E}
      \right) \rhd D_f
    \end{equation*}
    with \( (\mu, X) \in F^j \) for some row
    number \( j \) and we finally define
    \begin{equation*}
      \left(
        \mvec{\rho}, \pattMat{P}{\mvec{N}}{\mvec{C}}{\mvec{r}}, n,
        \mathcal{E}
      \right) \rhd
      \trbinCl(D_s, (\mathcal{E}(\mu), X), D_f)
    \end{equation*}
  \item\label{enum:tc-rec-nl} A non linearity constraints can be
    enforced
    when the two terms involved in the constraint have been stored,
    that is, when there is a couple
    \( \{ \mu, \nu \} \) (\( \mu \ne \nu \)) such that
    \( \mathcal{E}(\mu) \) and \( \mathcal{E}(\nu) \) are defined
    and there is a row \( j \) such that \( \{ \mu, \nu \} \in N_j \).
    If it is the case, then compute \( \condsucc \), \( \condfail
    \) and compile them,
    \begin{equation*}
      \label{eq:treecomp-binctrees}
      \left(
        \mvec{\rho}, \condsucc\left(\{\mu, \nu\},
          \pattMat{P}{\mvec{N}}{\mvec{C}}{\mvec{r}} 
        \right), n,
        \mathcal{E}
        \right) \rhd D_s
    \end{equation*}
    \begin{equation*}
      \left(
        \mvec{\rho}, \condfail\left(\{\mu, \nu\},
          \pattMat{P}{\mvec{N}}{\mvec{C}}{\mvec{r}}
        \right), n,
        \mathcal{E}
      \right) \rhd D_f
    \end{equation*}
    and define
    \begin{equation*}
      \label{eq:treecomp-bincstr}
      \left(
        \mvec{\rho}, \pattMat{P}{\mvec{N}}{\mvec{C}}{\mvec{r}} , n,
        \mathcal{E}\right)
      \rhd
      \trbinNl(D_s, \{\mathcal{E}(i), \mathcal{E}(j)\}, D_f)
    \end{equation*}
  \end{enumerate}

  \item\label{enum:tc-swap}
    If column \( i \) contain either a symbol, an
    abstraction or a constraint, and each pattern vector of \( P \) is at
    least of length \( i \), then compile
  \( \left(\mvec{\mu}, \pattMat{P'}{\mvec{N}}{\mvec{F}}{\mvec{r}}, n,
    \mathcal{E}\right)
  \rhd D' \) where
  \( \mvec{\mu} = (\rho_i\, \rho_1\, \dots\, \rho_n) \) and
  \( P' \) is \( P \) with column \( i \) moved
  to the front; to build
  \begin{equation}
    \label{eq:treecomp-swap}
    \left(
      \mvec{\rho}, \pattMat{P}{\mvec{N}}{\mvec{C}}{\mvec{r}},
      n, \mathcal{E}
    \right) \rhd \trswap_i(D')
  \end{equation}
\end{enumerate}
\end{definition}

\begin{example}[\autoref{ex:match}, \ref{ex:tree} continued]
  We consider again the rewriting system used in
  \autoref{ex:match}. We start by computing the matrices:
  \[ \pattMat{P}{\emptyset}{\emptyset}{\mvec{r}} \mdefn
    \pattMat{%
      \begin{bmatrix}
        \rws{c}\ (\rws{c}\ \patt) & \rws{a}\\
        \patt & \rws{b}
      \end{bmatrix}
    }{%
      \begin{bmatrix}
        \emptyset \\ \emptyset
      \end{bmatrix}
    }{%
      \begin{bmatrix}
        \emptyset \\ \emptyset
      \end{bmatrix}
    }{%
      \begin{bmatrix}
        \patt[x] \\ \patt[x]
      \end{bmatrix}
    }.
  \]
  \begin{enumerate}
  \item
    We saw that it is better to start examining the second argument,
    so we start by swapping columns of the matrix,
  \( P' =
  \begin{bmatrix}
    \rws{a} & \rws{c}\ (\rws{c}\ \patt)\\
    \rws{b} & \patt
  \end{bmatrix} \),
  define \( D \) such that
  \( (2\, 1), \pattMat{P'}{\emptyset}{\emptyset}{\mvec{r}}, 0, \emptyset \rhd D \).
  and we thus have
  \[ ((1\, 2), \pattMat{P}{\emptyset}{\emptyset}{\mvec{r}}, 0, \emptyset)
    \rhd
    \trswap_2(D). \]

  \item
  To continue and compute \( D \),
  we can match on the symbols of the first column of \( P' \) with a
  \( \trswitch \) node. For this, we compute
  \begin{itemize}
  \item \( P_{\rws{a}} = \specialise(\rws{a}, 0, P') =
    \begin{bmatrix} \rws{c}\ (\rws{c}\ \patt) \end{bmatrix} \), and
  \item \( P_{\rws{b}} = \specialise(\rws{b}, 0, P') =
    \begin{bmatrix} \patt \end{bmatrix} \).
  \end{itemize}
  Then we compute \( D_{\rws{a}} \) and \( D_{\rws{b}} \) such that
  \( (2, \pattMat{P_{\rws{a}}}{\emptyset}{\emptyset}{\mvec{r}}, 0,
  \emptyset) \rhd D_{\rws{a}} \) and
  \( (2, \pattMat{P_{\rws{b}}}{\emptyset}{\emptyset}{\mvec{r}}, 0,
  \emptyset) \rhd D_{\rws{b}} \).
  The switch case list \( L \mdefn [(\rws{a}_0, D_{\rws{a}}),
  (\rws{b}_0, D_{\rws{b}})] \) can be defined and so the compilation
  step produces
  \[ ((2\, 1), \pattMat{P'}{\emptyset}{\emptyset}{\mvec{r}}, 0,
    \emptyset) \rhd \trswitch(L). \]

  \item
    Since \( P_{\rws{b}} \) contains only unconstrained variables, we
    are in the case \autoref{enum:tc-vars} and so we have
    \( (1, \pattMat{P_{\rws{b}}}{\emptyset}{\emptyset}{\patt[x]}, 0,
    \emptyset)
    \rhd \trleaf(\patt[x]) \).

    \item
      A specialisation on \( P_{\rws{a}} \) with respect to $\rws{c}$ can be performed,
      let \( Q_{\rws{a}} \mdefn \specialise(\rws{c}, 0, P_{\rws{a}}) =
      \begin{bmatrix} \rws{c}\ \patt \end{bmatrix} \) and define \( E \)
      such that
      \( (1, \pattMat{Q_{\rws{a}}}{\emptyset}{\emptyset}{\patt[x]}, 0,
      \emptyset)
      \rhd E \). The compilation step
      produces
      \[ (1, \pattMat{P_{\rws{a}}}{\emptyset}{\emptyset}{\patt[x]}, 0,
        \emptyset)
        \rhd \trswitch([(\rws{c}, E)]). \]
    \item
      Similarly, we can specialise \( Q_{\rws{a}} \) on \( \rws{c} \)
      yielding the matrix \( \begin{bmatrix} \patt \end{bmatrix} \)
      which compiles to \( \trleaf \). We thus have,
      \[ (1, \pattMat{Q_{\rws{a}}}{\emptyset}{\emptyset}{\patt[x]}, 0,
        \emptyset) \rhd \trswitch([(\rws{c}, \trleaf(\patt[x]))]). \]
\end{enumerate}
\end{example}

The soundness and completeness proofs for this
compilation process can be found in \cite{hondet19master}.

We have seen that at each compilation step, several options are
possible. The stack can be swapped with \( \trswap \) to orient the
filtering. If a constraint can be solved, either is is solved with a
\( \trbinNl \) or \( \trbinCl \) node, or a \( \trswitch \) can be
performed. These possibilities make the compilation process
undeterministic. Therefore, a given matrix can be compiled to
several decision trees. Maranget compares different heuristics
based on the shape of patterns as well as some more complex ones.
In \dedukti, since verifying constraints can involve non trivial
operations (non-linearity and variable occurrence tests), constraint
checking is postponed as much as possible.
Regarding \( \trswap \), we process in priority columns that have
many constructors and few constraints.


\section{Results}
\label{sec-results}

This section compares the performance of the new rewriting engine with
previous implementations of \dedukti{}, and other tools as well.

\subsection{Hand-written examples}

We consider 3 different implementations of \dedukti{}:
\begin{itemize}
\item \toolstyle{Dedukti2.6} is the latest official release of
  \dedukti{} available on {\tt opam}. Its matching algorithm also
  implements decision trees but non-linearity and variable-occurrence
  constraints are not integrated in decision trees. Its
  implementation, primarily due to Ronan Saillard
  \cite{saillard15phd}, is available on
  \url{https://github.com/Deducteam/dedukti}.
  
\item \toolstyle{Lambdapi1.0} is an alternative implementation of
  \dedukti{} due to Rodolphe Lepigre \cite{lepigre18lfmtp}. It
  implements a naive algorithm for matching. It is available on
  \url{https://github.com/rlepigre/lambdapi/tree/fix\_ho}.

\item \toolstyle{Dedukti3.0} is our new implementation of \dedukti{}.
  It adds to \toolstyle{Lambdapi1.0} the decision
  trees described in this paper. It is available on
  \url{https://github.com/Deducteam/lambdapi}.
\end{itemize}

The git repository \url{https://github.com/deducteam/libraries}
contains several hand-written \dedukti{} examples, including a Sudoku
solver with 3 examples labelled easy, medium and hard respectively,
and a DPLL-based SAT solver to decide the satisfiability of
propositional logic formulae in conjunctive normal form with two
example files:
\begin{itemize}\itemsep=0mm
\item\texttt{2\_ex}
    contains a function that when given a integer \( n \),
    produces \( n \) literals named \( v_n \) and the formula
    \( p(0) = v_0 \land
    \bigwedge_{k=1}^{n} (p(k) = p(k - 1) \land (v_{k-1} \ne v_{k})) \)
\item\texttt{ok\_50x80}
    contains a formula with 50 literals and 80 clauses of the form
    \( \neg x \lor \neg y \lor \neg z \).
\end{itemize}
Because of the nature of the problems,
they require a substantial amount of rewriting steps to be solved.

\begin{table}
  \centering
  \caption{Time needed to solve Sudoku and SAT formulae in
    seconds.}\label{tab:sud-dpll}
  \begin{tabular}{l r r r r r r}
    \toprule
    & \multicolumn{3}{c}{Sudoku} & \multicolumn{2}{c}{DPLL-SAT}\\
    & easy & med & hard & \texttt{2\_ex} & \texttt{ok\_50x80}\\
    \midrule
    \toolstyle{Dedukti2.6} & 0.7 & 7.7 & 8 min 43 & 2 & 10\\
    \toolstyle{Lambdapi1.0} & 1.2 & 13 & 16 min 2 & 1.6 & 10\\
    \toolstyle{Dedukti3.0} & 0.5 & 5.2 & 5 min 15 & 0.2 & 2\\
    \bottomrule
  \end{tabular}
\end{table}

\autoref{tab:sud-dpll} shows the performance of each tool on these examples.
Using decision trees increases significantly performance on Sudoku
since \toolstyle{Lambdapi1.0} is twice as slow as
\toolstyle{Dedukti2.6} which is slower than \toolstyle{Dedukti3.0}.
SAT problems confirm that \toolstyle{Dedukti3} is more efficient than
\toolstyle{Lambdapi1.0} and \toolstyle{Dedukti2}.

More benchmarks are described in \cite{hondet19master}.

\subsection{Rewriting Engine Competition (REC)}

The Rewriting Engine
Competition\footnote{\url{http://rec.gforge.inria.fr}} (REC), first organized in 2009, aims to
compare rewriting engines.
F.~Duràn and H.~Garavel revived the competition in 2018, and another study has been
done in 2019 \cite{Durn:REC19}. There are 14 rewriting engines
tested, among which Haskell's \toolstyle{GHC} and \toolstyle{OCaml}.

REC problems are written in a specific \toolstyle{REC} syntax which is
then translated into one of the 14 target languages with
\toolstyle{Awk} scripts.
To use \toolstyle{REC} benchmarks with \dedukti{},
a translation from \toolstyle{Haskell} benchmarks to \dedukti{}
has been
implemented\footnote{file \texttt{tools/rec\_to\_lp/rec\_hs\_to\_lp.awk}
  available from revision e8388b73 (published on May 12, 2020)}.

Our rewriting
engine\footnote{with revision a0009fdaa (published on Jan. 10, 2020)}
has been compared on the problems that do not use
conditional rewriting with \toolstyle{OCaml} and \toolstyle{Haskell}.
For each language,
we have measured both the interpretation time
with \texttt{ocaml} and \texttt{runghc}, and the compiling and running
time with \texttt{ocamlopt} and \texttt{ghc}.
The results are in \autoref{tab:rec}.

We can divide our observations on classes of problems.
There are 43 problems, among which 22 are solved in less that one
second by at least two other solvers than \dedukti\
(the first group of the table).
On these problems, our rewriting engine is in average 4 times faster
than the median of the other rewriting engines.
The second group contains problems on which
no other tool than \dedukti\ needs more than ten seconds.
On this group, \dedukti\ is in average 10 times slower than the median of the
other tools.
However, \dedukti\ performs better than
interpreted \toolstyle{OCaml} on \texttt{add8}
and better than compiled \toolstyle{OCaml} on\texttt{benchtree10}.
On the last group, \dedukti\ is in average 60 times slower than other engines.
Interestingly \texttt{ocamlopt} has more memory overflows than \dedukti{}
(seven against four).

\begin{table}
  \centering
  \caption{Performance on REC benchmark in seconds.
    {\footnotesize N/A is for out of memory. T/O is for timeout (30 minutes).
      The last line indicates that on the \texttt{langton7} problem,
      \dedukti{} ran out of memory,
      the command \texttt{runghc langton7.hs} took 533.2 seconds to finish,
      \texttt{ocaml langton7.ml} took 101.7 seconds,
      {\tt ghc langton7.hs \&\& ./langton7} took 66 seconds and
      {\tt ocamlopt langton7.ml \&\& ./a.out} took 39.6
      seconds.}}\label{tab:rec}
  \begin{tabular}{l r r r r r}
  \toprule
  & \toolstyle{Dedukti} & \texttt{runghc} & \texttt{ocaml} & \texttt{ghc} & \texttt{ocamlopt}\\
  \midrule
revelt & 0.026 & 0.517 & 0.065 & 0.271 & 0.168\\
check1 & 0.030 & 0.417 & 0.065 & 0.270 & 0.170\\
calls & 0.017 & 0.416 & 0.065 & 0.271 & 0.168\\
check2 & 0.033 & 0.466 & 0.065 & 0.271 & 0.168\\
garbagecollection & 0.033 & 0.416 & 0.115 & 0.271 & 0.170\\
fibonacci05 & 0.033 & 0.416 & 0.065 & 0.271 & 0.168\\
soundnessofparallelengines & 0.033 & 0.467 & 0.065 & 0.271 & 0.168\\
factorial5 & 0.033 & 0.416 & 0.066 & 0.271 & 0.168\\
empty & 0.033 & 0.416 & 0.065 & 0.271 & 0.170\\
revnat100 & 0.064 & 0.516 & 0.115 & 0.276 & 0.170\\
factorial6 & 0.065 & 0.467 & 0.066 & 0.271 & 0.169\\
tautologyhard & 0.065 & 0.716 & 0.165 & 0.271 & 0.221\\
fibonacci18 & 0.115 & 0.466 & 0.065 & 0.275 & 0.170\\
fibonacci21 & 0.166 & 0.566 & 0.068 & 0.178 & 0.174\\
benchexpr10 & 0.165 & 0.667 & 0.266 & 0.275 & 0.221\\
benchsym10 & 0.165 & 0.667 & 0.266 & 0.275 & 0.221\\
natlist & 0.165 & 0.868 & 0.266 & 0.275 & 0.220\\
fibonacci19 & 0.168 & 0.517 & 0.065 & 0.174 & 0.170\\
factorial7 & 0.215 & 0.467 & 0.065 & 0.275 & 0.170\\
fibonacci20 & 0.265 & 0.567 & 0.065 & 0.283 & 0.174\\
permutations6 & 0.366 & 0.868 & 0.115 & 0.299 & 0.182\\
factorial8 & 1.467 & 0.817 & 0.115 & 0.299 & 0.183\\
\midrule
revnat1000 & 3.300 & 3.578 & 0.266 & 0.482 & 0.281\\
benchtree10 & 3.476 & 0.667 & 126.027 & 0.275 & 11.546\\
factorial9 & N/A & 2.974 & N/A & 0.532 & 0.282\\
permutations7 & 6.376 & 3.276 & 0.416 & 0.531 & 0.331\\
add8 & 10.899 & N/A & 12.605 & 0.232 & N/A\\
benchsym20 & 14.335 & 6.581 & 1.067 & 0.734 & 0.381\\
benchexpr20 & 14.787 & 6.786 & 1.417 & 0.932 & 0.983\\
\midrule
mul16 & T/O & 11.154 & 6.640 & 0.735 & N/A\\
add32 & T/O & 19.714 & 8.640 & 0.331 & N/A\\
benchtree20 & N/A & 21.333 & T/O & 3.301 & T/O\\
mul32 & T/O & 27.562 & 10.444 & 1.638 & N/A\\
benchsym22 & 54.491 & 23.534 & 2.727 & 1.435 & 0.833\\
benchexpr22 & 52.995 & 24.492 & 3.979 & 2.092 & 2.337\\
add16 & 79.472 & 22.697 & 7.140 & 0.303 & N/A\\
mul8 & 167.500 & 4.479 & 3.771 & 0.331 & N/A\\
omul32 & T/O & 49.863 & 25.251 & 1.885 & N/A\\
benchtree22 & N/A & 101.176 & T/O & 10.047 & T/O\\
revnat10000 & 400 & 244.459 & 6.987 & 19.241 & 4.092\\
omul8 & 797.406 & 5.430 & 3.971 & 0.381 & N/A\\
langton6 & N/A & 377.208 & 75.709 & 38.437 & 24.463\\
langton7 & N/A & 533.197 & 101.695 & 66.093 & 39.640\\
\bottomrule
\end{tabular}
\vspace*{2mm}
\end{table}


\section{Conclusion \& Related Work}
\label{sec-conclu}

This article describes the implementation of the new rewriting engine
of \dedukti. It extends Maranget's techniques of decision trees used
in the OCaml compiler \cite{maranget08ml} to the class of non-linear
higher-order patterns used in Combinatory Reduction Systems (CRS)
\cite{klop93tcs}. We define the language of decision trees and how to
compile a set of rewrite rules into a decision tree. We finally
present some benchmarks showing good performances.

A similar algorithm had been implemented in \toolstyle{Dedukti2.6} by
Ronan Saillard \cite{saillard15phd}. However, Saillard's rewriting
engine used decision trees for first-order linear matching and handled
non-linearity and variable-occurrence constraints afterwards in a
naive way. In the new implementation, these constraints are fully
integrated in decision trees.

Other rewriting engine uses decision trees as well like CRSX, which is
a rewrite engine for an extension of Combinatory Reduction Systems
\cite{rose:LIPIcs:2011:3130}, and Maude under certain conditions
\cite{eker:1996:fast-matching}, but Maude considers first-order terms
only.

Other pattern-matching algorithm are also possible, in particular
using backtracking automata instead of trees, which allow
to have smaller data structures.
The interested reader can look at Prolog implementations or
\toolstyle{Egison} (see \cite{Egi:EgisonNP15} and, more particularly on
the question of pattern matching, \cite{Egi:NonlinearPM18}).

Further useful extensions would be interesting: conditional rewrite
rules (the REC database contains many files with conditional rewrite
rules) and matching modulo associativity and commutativity (AC).
Conditional rewriting could be implemented without too much difficulty
since it would consist in extending the constraints mechanism which is
modular. A prototype implementation of matching modulo AC has
already been developed for \toolstyle{Dedukti2.6} by Gaspard
F\'erey\footnote{revision 5990bc6c (published on Feb. 21, 2021)}
but performance is not very good yet. This new implementation could
provide a better basis to implement matching modulo AC.


\bibliography{fscd20}
\end{document}